\documentclass{article}
\usepackage{graphicx,url,hyperref}
\usepackage[british]{babel}
\usepackage[round]{natbib}
\usepackage[official]{eurosym}
\begin{document}

\title{How to Improve Portuguese Secondary Education}
\author{C. J. A. P. Martins$^{1,2}$\thanks{Carlos.Martins@astro.up.pt}}
\date{\small
    $^1$Centro de Astrof\'{\i}sica da Universidade do Porto, Rua das Estrelas, 4150-762 Porto, Portugal\\%
    $^2$Instituto de Astrof\'{\i}sica e Ci\^encias do Espa\c co, Universidade do Porto, Rua das Estrelas, 4150-762 Porto, Portugal\\[2ex]%
    Last edited on 20 January 2026
}
\maketitle

\section*{Abstract}    
I share some personal thoughts on the status of Portuguese secondary education in general, and of the physics part thereof in particular, drawn from several decades of experience of organizing training activities for students and school teachers, as well as several hundred visits to secondary schools and similar numbers of interviews of secondary school students (from Portugal and elsewhere). I offer various suggestions for improving and modernising our current system.

\section*{Introduction}

Throughout my three decades as a researcher in physics and astronomy, I have kept an active interest in the public understanding of science, and more specifically in teaching, education and outreach. Inter alia, I have given more than 300 outreach talks, the vast majority of which in more than 150 different Portuguese secondary schools. The interactions with these student audiences (as small as 15 and as large as 400) and the accompanying teachers often devolve into discussions about the school curriculum. I have also taught many certified teacher training courses (most of them certified for physics \& chemistry teachers, some also for biology \& geology and for maths teachers), I have created, and am the scientific director of the AstroCamp\footnote{\url{https://astrocamp.astro.up.pt/}} \citep{astrocamp1}, an international astrophysics summer school for secondary school students, and have also supervised various projects and school science clubs, and hosted internships for secondary school students at my research institution.

These activities, and the regular detailed interactions with students and teachers (and, less frequently, with school directors and others involved in the system) led me to develop an eagle’s eye view of the Portuguese secondary education system – one that is, unfortunately, increasingly worrying. I'm worried that a take-it-easy mentality is being encouraged, which disregards analysis and critical reasoning. I'm shocked by the return to scholastic methods that leave students clearly incapable of applying the knowledge they supposedly possess to new situations. And while there is a wide range of teachers’ pedagogical skills and motivation, it saddens me to see the more active and committed ones submerged in bureaucracy and without freedom of initiative. It's obvious that the massification of education in Portugal was achieved at the cost of a drastic reduction in the standards of demand – it suffices to compare recent national exams in mathematics or physics with those from, say 20 years ago. The price is the emergence of a generation with an alarming level of scientific illiteracy and innumeracy.

In what follows I offer some ideas for improving this situation. Some of these points pertain specifically to physics and related areas, in which I have technical expertise, but I place them in a broader context since a broader revision of the system architecture is evidently necessary.

\section*{The status quo}

The current Portuguese secondary school system (10th to 12th grades, nominally for 15 to 18 year olds) faces challenges on at least four aspects. The first is the curriculum. Its main bottlenecks are a sloppy mathematical background (e.g., derivatives and statistics concepts are introduced relatively late, and integrals not at all) and a poor classical vs modern physics balance. Absurdly, physics and chemistry are a single subject in both 10th and 11th grade, and only become separate ones in 12th grade, at which point they are optional, even for students in the science and technology area (one of the four available areas). So Portuguese students can enroll in a physics or chemistry university degree without taking either subject in 12th grade. This is especially applicable to physics, which is considered a difficult 12th grade subject (more precisely: one where achieving a good grade requires work), leading to the temptation of choosing more fluffy subjects. Indeed, in many schools, 12th grade physics is not offered at all, since doing so requires a minimum number of interested students. Those who are indeed interested often need to change school at this point.

School facilities are widely discrepant, and the differences are not simply between large wealthy cities and poor small towns. I know several examples of pairs of schools less than 1 km from each other with totally different facilities---as well as different internal rules and traditions for their use. Moreover, in recently built or refurbished schools there are some shocking design flaws in auditoriums and science labs, making one wonder if the responsible architects know how school facilities and labs are used. 

Unfortunately, teachers are also a bottleneck. To be clear, there's a wide range of technical and pedagogical skills, as well as motivation, among them – and specifically physics teachers, which are the ones I’m qualified to thoroughly assess. Some are truly excellent, but for others teaching was only a Plan B, and it shows. Most teachers are in their 50s or 60s, which among other things means that they have no coding skills (and no interest in acquiring them), on which more below. A system of certified teacher training courses exists, but the system is quite often abused, and most of these courses are merely perfunctory, providing no substantive training.

Finally, there is teaching-to-the test approach. Certainly this is not unique to Portugal, but the media seems increasingly obsessed with school rankings which are entirely without merit. In my experience, rankings tell you absolutely nothing useful, and this is especially so when it comes to the top students. The consequence is that schools don’t train students to learn, they only train them to do one particular type of exam, which is not even the type of exam students will find in university – at least if they enroll in physics, engineering or other substantive degrees. There's also an emerging trend of some ‘educators’ emphasising communication skills instead of knowledge, much like sophists emphasised rhetoric. This is a fatal system flaw: illiteracy should not be confused with charm, or authenticity with merit.

\section*{How to fix it: programming and projects}

Having identified some problems, I will now offer some ways to fix them. First and foremost, the science and technology branch of Portuguese secondary education must have a second core subject, in addition to mathematics, spanning all three years: scientific programming, using languages such as Python \citep{python}, Octave \citep{octave} or Julia \citep{julia}. Currently there’s an optional 12th grade subject, dubbed ‘Computing Applications’, which in a nutshell, is just plain stupid, unless your only goal is to get a lazy grade---or students have the good fortune of having a good teacher and everyone is willing to work for a year. My own experience of undergraduate and student supervision is that the best outcome predictor for physics and engineering physics undergraduate students is not their university entry grade: it’s whether or not they had good programming skills when they started university.

This will require a training effort, since the vast majority of Portuguese physics teachers (or other science teachers) can’t code, but it is an essential step. I would go so far as to say that physics teachers who can’t code should not be teaching secondary school physics, but must retrain or be replaced. In fact, given that physics and chemistry are a single subject in 10th and 11th grade, secondary school physics in Portuguese schools is mainly taught by teachers whose core training is in chemistry, which sometimes impacts the quality of the physics teaching.

A second point is for each student to develop a secondary school project – a school graduation thesis in a natural science subject (physics, chemistry, geology, or biology), according to the student’s choice, but also building upon the programming subject, and ideally drawing upon statistics and data analysis and modelling covered in mathematics. The experience of dealing with real, noisy data is a crucial one (and not just for future scientists), and in my view one of the tasks of modern secondary education is to provide an environment for a first such experience. This thesis also develops scientific writing and other skills, including autonomy and resilience---which are in extremely short supply. This thesis would be defined in 10th grade (following some background reading), developed in 11th grade (with experiments, data collection, etc.), and written and presented in 12th grade, and would be supervised by a school teacher in the most relevant area, ideally with a co-supervisor in academia, industry, or another real-world context. In Portugal it used to be said that universities were an ivory tower world, with no connection to the real one, but today it is secondary schools who are in that role.

Then there is the the issue of what is actually taught and how. Evidently physics and chemistry should be separate subjects from 10th grade, even if remaining optional in 12th grade. The content of both needs urgent reform, to cover, for example, the physics and chemistry of global warming. In my frequent visits to schools to talk about physics topics, this is by far the most requested topic, and also the one for which teachers most complain about lack of clear and rigorous information. (Naturally, it's a topic that also impacts biology and geology.) Additionally, this content revision can update some terminology in physics (and undoubtedly in other sciences) which was used in 1980s but is no longer used at university level, forcing junior undergraduates to relearn it.

\section*{Interlude: physics problems}

A further weakness of the Portuguese secondary education system is that not only there’s no astronomy subject (unlike several other countries), but there is almost no astronomy-related content in physics or any other secondary school subject. Such content, which did exist previously, has been gradually removed. For example, until a few years ago 10th grade chemistry started with the Big Bang model and some stellar physics, since primordial and stellar nucleosynthesis are behind the formation of a good fraction of the elements of the periodic table (the real topic of interest). This is unfortunate given the inspirational potential of astronomy, and is likely due to the fact that the Portuguese Physics Society is usually involved in curriculum revisions, while its Astronomy counterpart is not.

There are disconnects, e.g. logarithms are used in 11th grade physics and chemistry but they are only defined in 12th grade mathematics, presumably because those who created each subject’s curriculum never considered the others. And there are some errors and inaccuracies, which are more difficult to accept. The one I find most shocking, given my field or research, is that the 11th grade physics syllabus confuses the cosmological expansion with the Doppler effect.

The physics of the Doppler effect is based on the relative velocity between a source and an observer. Cosmologically, this would occur in a Newtonian universe, where space and time would be absolute and immutable, populated by a grid of identically calibrated rulers and synchronized clocks. The motion of galaxies in that absolute space, moving away from the observer, would then, through the Doppler effect, redshift the wavelengths, and these would remain constant along the way until we measure them. The idea is simple, intuitive, and easy to explain. It is also, physically, wrong. This is not how the Universe works, and this supposed coordinate grid, which is drawn in kinematics exercises, does not exist. What's more, it is demonstrably wrong, and I know quite a few secondary school students who have experimentally demonstrated this themselves, by building a Wilson cloud chamber and measuring the variation of the muon flux with altitude.

In the cosmological expansion there are no relative velocities: ignoring peculiar velocities due to local gravitational fields, galaxies are at rest when they emit photons. The expansion of the universe is an expansion of space itself (more strictly, of the spatial component of spacetime), which stretches all comoving distances, and in particular, gradually redshifts wavelengths until they reach us. For each galaxy, the measured redshift is generally different from that which would result from the Doppler effect.

One possible reason for the conceptual confusion, sometimes mentioned by teachers who understand the issue, is that for low distances and redshifts, the mathematical formula describing this cosmological expansion coincides with the Doppler effect formula. That is true, but it means nothing. Mathematics is not physics: mathematics is a science, but it is not a natural science, and the fact that the same mathematical formula is applicable in two different physical contexts does not in any way imply that its physical interpretation in both cases is the same. This is why it is impossible to learn physics by trying to memorize mathematical formulas. It’s also a good example of a missed opportunity in the education system: in my personal experience, 11th grade students can understand this difference if it is adequately explained to them \citep{astrocamp2}.

The relationship between General Relativity and Newtonian physics is itself a good example of this conceptual difference. There is a mathematical relationship between the two: by taking a specific limit of General Relativity (small curvature, low velocity, etc.), the equations of Newtonian physics can be recovered. However, the physical interpretations of the two theories are not only different but mutually incompatible, and again, the Newtonian interpretation is demonstrably wrong. Here it is considerably more difficult for high school students to experimentally show it, but it is not impossible: one way to do it is to carry out a modern version of the Eddington Experiment \citep{Bruns}, as the AstroCamp students and several other international groups will be doing in 2026-2027.

\section*{How to fix it: exams and teachers}

In my lifetime, the structure of national exams in Portugal has changed a few times, sometimes for the better and sometimes for the worse, as in the most recent one. At the moment, some national exams (like the Physics \& Chemistry one) are done at the end of  11th grade, and only Mathematics and Portuguese are mandatory at the end of 12th grade. That’s simply poor judgment, considering the growing immaturity of students of constant chronological age. All national exams, and certainly the Physics one, should be done at the end of 12th grade. It would be then straightforward to make 12th grade physics compulsory for any university physics or engineering degree, as simple common sense demands.

An objection sometimes raised is that students wouldn’t cope with the stress of multiple exams in a short time (which is an euphemism for saying that they are increasingly immature), but this is easy to solve, relying on the fact that in 12th grade the total number of contact hours in the academic year is about 2/3 of that in 10th and 11th grade. The solution is to have all classes in the 1st and 2nd terms (i.e., until the Easter break). Then the period from Easter to June/July would be free from classes, and students can prepare for exams and finish their thesis. Both the exams (say 4-5 per student) and the thesis presentation can be spread across June and July – a more than comfortable schedule for anyone aiming to do a university degree.

Since the exams are used for university admissions, they must be set by university lecturers or researchers, not by school teachers – who typically have decoupled from contemporary academia and are not qualified to set exams that meaningfully test the student’s preparation and skills vis a vis a particular academic discipline. This is one reason why many students enter university with a very wrong idea of what the degree will be like, and explains why in competitive courses like physics there’s a significant early drop-out rate: students suddenly realise that the degree is very different from their naïve expectations.

Personally, and based on many years of experience, I’m also a strong advocate of entrance interviews for some university degrees (especially those where grades alone can’t predict exit outcomes, such as physics), but I concede that this would entail larger logistics difficulties than the previous points. Still, in Portugal, this would mitigate the obvious differences in how generous public and private schools are in awarding high grades, which, given our numerus clausus system, often causes unfair situations.

Another desirable (but possibly difficult to implement) feature would be to reward the students obtaining the top results in the national exams with summer internships at universities, research labs, tech companies, or other suitable external institutions, an antidote to the aforementioned ivory tower issue. One could collect a pool of available internships to be matched, by their hosts, with the eligible students’ expressions of interest. Several other national and university-level programs already have this general format. Some of these internships could even include a small stipend.

And what will teachers do in this third term? That’s the easiest part---they will have some weekly office hours where they clarify student doubts as they prepare for exams, but they will have a significantly reduced teaching load, and that can use that time to take certified  meaningful teacher training courses. Only one further tweak is needed: it’s not uncommon in Portuguese schools for students to have three different physics teachers in three years (if they take 12th grade physics, that is), simply because teachers find it less of a hassle to always teach the same school grade. Instead, each student cohort should have the same physics teacher (or any other subject’s teacher) in all three years. In addition to the obvious mutual advantages one ensures that each teacher will have one ‘re-training term’ every three years, ensuring a better prepared, more rested and more motivated teacher cohort.

\section*{Outlook: Impossible, you say?}

While I have seen some schoolteachers in Portugal being shocked (or, more accurately, scared) by some of the above measures, I must point out that none of them is entirely new---versions of each of them exist, mutatis mutandis, and work successfully in other countries. As far as I know there's no place where all of them exist simultaneously, but they all exist. They've been implemented, and anecdotal evidence from multiple AstroCamp students suggests that they work. There’s no reason---not even cost---why they can’t work together, and indeed there are clear synergies between some of them.

Clearly, they are not impossible. They may be difficult to implement, but the difficulty lies only in the system’s bureaucratic inertia and close-mindedness. It might take a few years, but they're entirely doable. In some cases, such as programming, my experience with international students indicates that Portugal is already more than a decade behind other countries. The longer we wait to modernise the system, the longer will be the cost for the next generation, and for society as a whole---and I’m not talking about economic cost. 

\section*{Acknowledgements}

{\small This work was financed by Portuguese funds through FCT (Funda\c c\~ao para a Ci\^encia e a Tecnologia) in the framework of the project 2022.04048.PTDC (Phi in the Sky, DOI 10.54499/2022.04048.PTDC). CJM also acknowledges FCT and POCH/FSE (EC) support through the Investigador FCT Contract 2021.01214.CEECIND/CP1658/CT0001 (DOI 10.54499/2021.01214.CEECIND/CP1658/CT0001).}

\bibliographystyle{plainnat}
\bibliography{main}
\end{document}